\documentclass[pra,superscriptaddress,,nofootinbib]{revtex4}
\usepackage{graphicx}
\usepackage{amsmath}
\usepackage{amssymb}
\usepackage{float}
\usepackage{algorithm}

\floatname{algorithm}{Protocol}
\usepackage{bm}
\usepackage{bbm}

\usepackage{color}
\def\proof{\noindent Proof:\\}

\def\EQ#1{\begin{eqnarray}#1\end{eqnarray}}
\newcommand{\djj}{d\kern-0.4em\char"16\kern-0.1em}

\newtheorem{prop}{Proposition}\def\PRO{\begin{prop}}\def\ORP{\end{prop}}
\newtheorem{coro}{Corollary}\def\COR{\begin{coro}}\def\ROC{\end{coro}}
\newtheorem{theo}{Theorem}\def\TH{\begin{theo}}\def\HT{\end{theo}}
\def\TH{\begin{theo}}\def\HT{\end{theo}}
\newtheorem{defi}[prop]{Definition}\def\DE{\begin{defi}}\def\ED{\end{defi}}
\newtheorem{lemme}[prop]{Lemma}\def\LE{\begin{lemme}}\def\EL{\end{lemme}}
\def\one{\mathbb{I}}
\def\hil{\mathcal{H}}

\def\ket#1{\left| #1 \right\rangle}
\def\bra#1{\left\langle #1 \right|}

\def\qed{$\blacksquare$}
\begin{document}
\title{Minimum-cost quantum measurements for quantum information}
\author{Petros Wallden}
\email{petros.wallden@hw.ac.uk}
\affiliation{SUPA, Institute of Photonics and Quantum Sciences, School of Engineering and Physical Sciences, Heriot-Watt University, Edinburgh EH14 1AS, UK}
\affiliation{Physics Department, University of Athens, Panepistimiopolis 157-71, Ilisia Athens, Greece}
\author{Vedran Dunjko}
\affiliation{Now at: Institute for Quantum Optics and Quantum Information,
Austrian Academy of Sciences, Technikerstr. 21A, A-6020 Innsbruck, Austria}
\affiliation{SUPA, Institute of Photonics and Quantum Sciences, School of Engineering and Physical Sciences, Heriot-Watt University, Edinburgh EH14 1AS, UK}
\affiliation{School of Informatics, Informatics Forum, University of Edinburgh, 10 Crichton Street, Edinburgh EH8 9AB, UK.}
\affiliation{Laboratory of Evolutionary Genetics, Division of Molecular Biology, Ru\djj er Bo\v{s}kovi\'{c} Institute, Bijeni\v{c}ka cesta 54, 10000 Zagreb, Croatia}

\author{Erika Andersson}
\affiliation{SUPA, Institute of Photonics and Quantum Sciences, School of Engineering and Physical Sciences, Heriot-Watt University, Edinburgh EH14 1AS, UK}

\begin{abstract}
Knowing about optimal quantum measurements 
is important for many applications in quantum information and quantum communication. However, deriving optimal quantum measurements is often difficult.
We present a collection of results for minimum-cost quantum measurements, and give examples of how they can be used.
Among other results, we show that a minimum-cost measurement for a set of given pure states is formally equivalent to a minimum-error measurement for mixed states 
of those same pure states. For pure symmetric states it turns out that 
for a certain class of cost matrices, the minimum-cost measurement is the square-root measurement. That is, the optimal minimum-cost measurement is in this case the same as the minimum-error measurement. 
Finally, we consider 
sequences of individual ``local" systems, and examine when the global minimum-cost measurement is a sequence of 
optimal local measurements. 
We  also a consider an example where the global minimum-cost measurement is, perhaps counter-intuitively,  not a 
sequence of local measurements, and discuss how this is related to related to the Pusey-Barrett-Rudolph argument for the nature of the wave function. 
\end{abstract}

\maketitle

\section{Introduction}


The problem of finding optimal quantum measurements which decode classical information stored in quantum states, with various optimization criteria, has been studied since the very beginnings of quantum information theory~\cite{Helstrom}. 
A common scenario is 
minimum-error measurements. Here, 
given a known ensemble of quantum states $\{\rho_i,\eta_i \}_i$, where  $\eta_i$ is the probability with which the state $\rho_i$ appears, the task is to find a measurement which minimizes the average error probability in the result. 
Somewhat more generally, different types of error in the result can carry different costs according to a so-called cost matrix. The measurement which minimizes the average cost is then called the minimum-cost measurement.
In a quantum communication situation, classical information $i$ could first be encoded into a quantum state $\rho_i$, after which one may want to decode it back to classical information via a quantum measurement.
For example, finding relevant optimal figures of merit 
often  plays an important role in security proofs of quantum cryptographic protocols, where an adversary tries to obtain information about a quantum state. 
Optimal so-called generalised quantum measurements are certainly not only of theoretical interest, but have also been experimentally realized on photons, see for example~\cite{Clarke}, on NV centres~\cite{NVPRL}, and could be realized on trapped ions or atoms with existing experimental means~\cite{atomPOM}.

Finding optimal quantum measurements is in general hard. Optimal strategies have been obtained for some special cases, with various assumptions on the initial states. For minimum-error measurements, for instance, the input states usually have to possess some kind of symmetry~\cite{Helstrom, mult_sym, mirror_sym, Chou_Hsu, Nakahira}. 
An exception is the minimum-error measurement for arbitrary pure qubit states, occurring with uniform probability, which was obtained by Hunter~\cite{min_err_qubit}. 
A general geometric structure of the minimum-error problem was given only recently~\cite{geom_gener}.
Minimum-cost settings have been much less studied~\cite{Helstrom, ErikaImperfect}.

In this paper, we study both minimum-error and minimum-cost measurements, and establish a link between minimum-cost measurements for pure states and minimum-error measurements for mixed states. We then apply the general results we obtain to 
symmetric states, 
and their natural generalization, states which are sequences of
(that is, tensor products of) symmetric states.
Symmetric states are ubiquitous in quantum information. 
Quantum key distribution (QKD) using the BB84 protocol \cite{BB84} or coherent states \cite{barbosa-2002,Sych_Leuchs}, universal blind quantum computing (UBQC) \cite{UBQC} and quantum digital signatures (QDS) \cite{QDS,Expr,QDS_VPE,QDS_Exp}, for instance, use trains of independent symmetric states, giving rise to a tensor product structure. Optimal measurements on whole trains of states, versus measurements on individual elements, are analogous to individual and collective/coherent attacks in QKD.

The outline of this paper is as follows.
We begin by proving some general results concerning minimum-cost measurements, and establish a formal equivalence between minimum-cost measurements for pure states and minimum-error measurements for mixed states.
Following this, we focus on the minimum-cost problem of the so-called symmetric states, for both mixed and pure states.
Finally, we explore the minimum-cost problem for states 
which are tensor products of individual (local) 
states, motivated by situations which often appear in quantum cryptographic protocols. We analyse when the local measurements are the minimum-cost, give example that the minimum-cost measurement is global and highlight a connection with the Pusey, Barrett and Rudolph (PBR) argument for the nature of the wave function~\cite{PBR} and quantum state elimination measurements~\cite{stevebook,OppenUSE}. We conclude with a brief discussion.




\section{\label{Section-general-results-cost}General results for minimum-cost measurements}

Suppose that some quantum states $\rho_i$ each occur with probability $\eta_i$, and that we are making a quantum measurement described by the measurement operators $\Pi_j$. We will denote the measurement by $\Pi$, and also define $B_{i,j}(\Pi)=Tr(\Pi_j \rho_i)$ as the probability to obtain result $j$ given that the state was $\rho_i$. Because probabilities have to be positive, it follows that the operators $\Pi_i$ have to be positive semi-definite.  Also, since probabilities for all possible outcomes (including not obtaining a result, if this may happen) should sum to one, it holds that $\sum_i \Pi_i=\one$.

Further, suppose that obtaining result $j$ when the state was $\rho_i$ carries a cost $C_{i,j}$. The average cost of the measurement $\Pi = \{\Pi_k\}_k$, with respect to the (real) cost matrix $C = [C_{i,j}]$ is denoted $\bar C(\Pi)$ and is given by
\EQ{\bar C(\Pi)= \sum_{i,j}\eta_i C_{i,j}Tr(\Pi_j \rho_i).}
The minimum cost is obtained by minimizing this average cost 
over all possible POVM's $\{\Pi_i\}$,
\EQ{\bar C_{min}=\min_{\{\Pi\}} \bar C(\Pi).}
It is well established~\cite{Helstrom} that a minimum-cost measurement is optimal if and only if the following criteria are met:

\begin{enumerate}

\item $\Gamma=\sum_j\Pi_jW_j=\sum_jW_j\Pi_j$ for $W_j=\sum_i\eta_i C_{i,j}\rho_i$.

\item $\Gamma=\Gamma^\dagger$.

\item $\Pi_j(W_j-\Gamma)=(W_j-\Gamma)\Pi_j=0$ for all $j$.

\item $(W_j-\Gamma)$ is positive semidefinite for all $j$.

\end{enumerate}
It can be shown that the three first conditions are equivalent to
\EQ{\Pi_i(W_i-W_j)\Pi_j=0.} This form of the conditions was first derived by Holevo \cite{Holevo} and Yuen et al. \cite{Yuen} independently.
We will refer to the criteria above, as is usually done, as the Helstrom criteria.

For minimum-cost measurements we can prove the following general properties, which we first give informally. Keeping the states $\rho_i$ and prior probabilities $\eta_i$ the same,
\begin{enumerate}
\item The optimal measurement remains the same 
if the same column is added to or subtracted from each of the columns of the cost matrix. 
This means that the costs associated with different outcomes, for the same prior state $\rho_i$, all shift by the same amount. The average cost will also shift by a fixed amount.
\item The average minimal cost is superadditive with respect to the cost matrix. This means that the sum of the optimal minimal costs for some cost matrices $C^1,\ldots, C^n$ is lower than the minimal cost 
for the cost matrix $\sum_{k=1}^{n}C^k$.
\item  Increasing (decreasing) each entry of the cost matrix by a varying amount increases (decreases) the optimal minimum cost of the problem. In other words, the minimum cost 
is monotone under the point-wise partial order of the cost matrices.
\end{enumerate}

A special class of minimum-cost problems is the well-studied minimum-error problem. In the minimum-error problem the task is to, given some fixed set of states with some prior probabilities, find the measurement (and the ensuing success probability) which, on average, minimizes the probability of 
an error in the result. 
It is easy to see that this is a special class of minimum-cost problems, for a cost matrix with elements $C_{i,j} =A - \delta_{i,j}$ for any (real) constant $A$. If we choose $A=1$, then the minimum cost $\bar C_{min}$ is the minimum-error probability.
At the end of this section, we will  show that there is an additional 
one-to-one correspondence between minimum-error measurements on mixed states and minimum-cost measurements for pure states.

Next, we will formally state and  prove the above 
claims for minimum-cost measurements. 
In this paper, whenever there is addition or subtraction in matrix indices, this is understood as modular addition or subtraction. 
For example, if $A$ is an $N\times M$ matrix, then $A_{i+N,j+M}=A_{i,j}$.

\LE\label{constant-row}
Assume a minimum-cost problem with the cost matrix with elements $C_{i,j}$, where the 
states $\rho_i$ appear with the frequencies $\eta_i$. If we add (subtract) a \emph{constant-row cost matrix} with elements $C^r_{i,j} = C^r_i$, to (from) the original cost matrix, i.e. $C^t_{i,j}=C_{i,j}\pm C^r_{i,j}$, then the following two properties hold. \begin{itemize}
\item[(a)] The measurement that gives the minimum cost for the problem with 
the cost matrix $C^t_{i,j}$ also gives the minimum cost for $C_{i,j}$. That is, the measurement that gives the minimum cost is not altered.
\item[(b)] The minimum cost of $C^t$ is equal to the minimum cost of $C$, shifted by the cost of the constant row matrix $\bar C^r=\sum_i \eta_i C^r_i$.
\end{itemize}
\EL

\proof
First note that a cost matrix with fixed elements in each row (a \emph{constant-row matrix}), i.e. $C_{i,j}=C_{i,j+k}=c_i\forall k$, gives the same cost for \emph{every} measurement, and this cost is equal to $\bar C^r=\sum_i \eta_i c_i$.
This follows from 
\EQ{\bar C^r(\Pi)=\sum_{i,j}\eta_i C^r_{i,j}Tr(\Pi_j\rho_i)=\sum_i \eta_i c_i Tr((\sum_j \Pi_j)\rho_i)=\sum_i\eta_i c_i=\bar C^r.}
Therefore, all measurements are optimal for such a minimum-cost problem.
For the situation in this lemma, the total cost is given by
\EQ{\bar C^t(\Pi)&=&\sum_{i,j}\eta_i (C_{i,j}\pm C^r_{i,j})Tr(\Pi_j \rho_i)=\sum_{i,j}\eta_i C_{i,j}Tr(\Pi_j \rho_i)\pm\sum_i \eta_i c_i\nonumber\\&=&\bar C(\Pi)\pm\sum_i \eta_i c_i=\bar C(\Pi)\pm \bar C^r.}
Now it is easy to see that our lemma holds as the second (additive) term on the rightmost side of the equation above is independent of the measurement $\Pi$.
Thus the changed cost matrix $C_{i,j}^t$ yields the same optimal measurement, with the minimum shifted by $\sum_i \eta_i c_i=\bar C^r$. \qed

\LE\label{sum of cost matrices}
Let $C$ be a cost matrix such that $C_{i,j}=\sum_k C^k_{i,j}$, for some individual cost matrices $C^k, k=1, \ldots n$. Then the minimum cost induced by the cost matrix $C$ is bounded from below by the sum of the individual minimum costs induced by the individual cost matrices appearing in the sum, i.e. $\bar C_{min}\geq \sum_k \bar C^k_{min}.$
\EL

\proof
For any measurement $\Pi$ it holds that
\EQ{\bar C(\Pi)=\sum_k \bar C^k(\Pi).}
Suppose that the measurement $\Pi'$ gives the minimum cost for the total cost matrix, and that the measurements $\Pi^k$ give the minimum costs for the cost matrices $C^k$, respectively. We then have
\EQ{\bar C_{min}=\bar C(\Pi')=\sum_k \bar C^k (\Pi')\geq \sum_k \bar C^k (\Pi^k)=\sum_k \bar C^k_{min}.}
\qed


\LE
Assume that we have a cost matrix $C=[C_{i,j}]$, and an element-wise smaller cost matrix $C^l = [C^l_{i,j}]$, with $C^l_{i,j}\leq C_{i,j}$ for all $i,j$, and an element-wise larger cost matrix $C^{u} = [C^{u}_{i,j}]$ with $C^u_{i,j}\geq C_{i,j}$ for all $i,j$. Then the minimum cost 
induced by the 
cost matrix $C$ is bounded from below by the minimum cost induced by $C^l$ and from above by the minimum cost of $C^u$. In other words, \EQ{\bar C^l_{min}\leq \bar C_{min}\leq \bar C^u_{min}.}
\EL
\proof
We can write $C_{i,j}=C^l_{i,j}+C^s_{i,j}$, where $C^s$ is a strictly positive cost matrix. If the cost matrix has only non-negative real elements then for any measurement $\Pi$ we have that  $\bar C^s(\Pi)\geq 0$. From Lemma \ref{sum of cost matrices} it follows that

\EQ{\bar C_{min}=\min_\Pi \bar C(\Pi)\geq \min_\Pi \bar C^l(\Pi)+\min_\Pi \bar C^s(\Pi)\geq \min_\Pi \bar C^l(\Pi)=\bar C^l_{min}}
Similarly, by noting that $C_{i,j}+C^s_{i,j}=C^u_{i,j}$ for some positive cost matrix $C^s$ we conclude that $\bar C_{min}\leq \bar C^u_{min}$.
\qed

We will use these lemmas  in the remainder of this paper.

\section{\label{Section-error-cost-general}Minimum-error measurements of mixed states as minimum-cost measurements of pure states}

Here we point out an 
equivalence between minimum-error measurements for mixed states and minimum-cost measurements for pure states.
Using the results in this subsection, we will then in the next section provide analytic bounds on the minimum-error probabilities of a wide class of mixed states, and also, for some special cases, give analytical expressions for the minimum-error probability.

As we have noted, a minimum-error measurement is simply a minimum-cost measurement for distinguishing between the same set of states, with a cost matrix given by $C_{i,j}=1-\delta_{i,j}$.
Suppose that we are interested in the minimum-error problem where the input states are a collection of  $N$ mixed states $\{\rho_i\}$, appearing with respective frequencies $\{ \eta_i \}_i$, of the form
\EQ{\rho_i=\sum_m a_{i,m}\ket{\psi_m}\bra{\psi_m},\label{eq:mixedstates}}
where $a_{i,m}$ are $N\times N$ coefficients such that $\sum_m a_{i,m}=1$, and $\{\ket{\psi_1},\cdots,\ket{\psi_N} \}$ are $N$ pure states.
Then, the minimum-error measurement 
minimizes the expression
\EQ{P_{err}(\Pi)=C(\Pi)=\sum_{i,j}\eta_i (1-\delta_{i,j})Tr(\Pi_j\rho_i)=1-\sum_i\eta_iTr(\Pi_i\rho_i).}
This minimum-error problem for the $N$ mixed states in (\ref{eq:mixedstates}), occurring with prior probabilities $\eta_i$,  is equivalent to a minimum-cost problem for the $N$ equiprobable pure states $\{\ket{\psi_j} \}_j$, with 
the  cost matrix
\EQ{\label{cost_error}C_{m,i}=1-N\eta_ia_{i,m}}
(note the inverse order of indices in $C_{m,i}$). 
This can be seen from the following derivation,
 \begin{eqnarray}
P_{err}(\Pi)&=&1-\sum_i\eta_iTr(\Pi_i\rho_i)=1-\sum_i\eta_iTr(\Pi_i\sum_ma_{i,m}\ket{\psi_m}\bra{\psi_m})\nonumber\\
& = & 1/N\left(\sum_{i,m}Tr(\Pi_i\ket{\psi_m}\bra{\psi_m})-\sum_{i,m}N\eta_ia_{i,m}Tr(\Pi_i\ket{\psi_m}\bra{\psi_m})\right)\nonumber\\
& = & 1/N\sum_{i,m}C_{m,i}Tr(\Pi_i\ket{\psi_m}\bra{\psi_m})=C(\Pi),
\end{eqnarray}
 where $C(\Pi)$ is the cost of the measurement corresponding to the POVM $\{\Pi\}$ for the pure states $\{\ket{\psi_m}\}$ with equal prior probabilities $1/N$ and cost $C_{m,i}$ that is defined in Eq. (\ref{cost_error}). This shows that for \emph{any} measurement (POVM) $\{\Pi\}$, the cost for the considered pure states is the same as the error probability for the mixed states. 
 It follows that the minimum-cost measurement for the pure states will also be the minimum-error measurement for the mixed states, with prior probabilities as stated above.
 Another thing to note regarding Eq. (\ref{cost_error}) is that in the case where $a_{i,j}=\delta_{i,j}$, it reduces to the usual formula for a minimum-error measurement on pure states.


\section{Minimum-cost measurements for pure symmetric states}

In this section we will consider 
minimum-cost measurements for pure symmetric states. In the section following this one, we will use these results 
to obtain the minimum-error probability for certain classes of mixed states, in particular, for mixed states that are mixtures of pure symmetric states.
We will first consider the Square Root Measurement (SRM), which is known to be the minimum-error measurement for pure symmetric states. We will express the success probability of the SRM (that is, the minimum-error measurement)
as a function of the eigenvalues of the Gram matrix of the states we are considering. Following this, we will extend the
minimum-error problem to a minimum-cost problem and prove that for certain class of cost matrices, the SRM is the minimum-cost measurement.
We will then apply the results of the previous sections to provide bounds for the minimum cost in an example, for 
four symmetric coherent states with equal amplitude but different phases.

Let $U$ be a unitary such that $U^N = I \label{p1}$. We define $\ket{\psi_i} = U^i \ket{\psi_0}$ for some $\ket{\psi_0}$. The $N$ states $\{\ket{\psi_0},\cdots,\ket{\psi_{N-1}}\}$ are called {\it symmetric}, and we will call $U$ the \emph{symmetry unitary}. We furthermore assume 
that the prior probabilities for the states are equal i.e. $\eta_i=1/N$. We define 
\EQ{B_{i,j}=Tr(\Pi_j \rho_i)=\bra{\psi_i}\Pi_j\ket{\psi_i},}
which is 
the probability that outcome $j$ is obtained, using the measurement $\{\Pi\}$, if the state sent was $\rho_i$. We can then rewrite the cost as
\EQ{\bar C(\Pi)=1/N\sum_{i,j} B_{i,j}C_{i,j}}
where we have used $\eta_i=1/N$.

\subsection{SRM measurement of symmetric states}

The square root measurement is 
known to be the minimum-error measurement for many cases, such as for pure symmetric states \cite{Helstrom}, for pure multiply symmetric states \cite{mult_sym} and for a certain class of mixed states \cite{Chou_Hsu} where at least one state has strictly positive coefficients when written in the symmetry operator eigenbasis. 
In the present paper, we will show that this measurement is important for a much wider range of cases, involving minimum-cost measurements and minimum-error measurements for certain mixed states  (exact conditions will be given later). We will also show how it is possible to bound the minimum cost and minimum-error probabilities for even more cases. If we define
\EQ{\Phi=\sum_{i=0}^{N-1}\ket{\psi_i}\bra{\psi_i},\label{eq:phi}}
then the square root measurement is defined by
\EQ{\Pi_j=\Phi^{-1/2}\ket{\psi_j}\bra{\psi_j}\Phi^{-1/2}=\ket{\phi_j}\bra{\phi_j}}
where
\EQ{\label{SRM-states}\ket{\phi_j}=\Phi^{-1/2}\ket{\psi_j}.}

The Gram matrix of the states we are trying to distinguish between 
is defined as
\EQ{
G_{i,j} := \left[ \bra{\psi_i} \psi_j \rangle    \right]_{i,j} = \left[ \bra{\psi_0} (U^{i})^\dagger U^{j}| \psi_0 \rangle    \right]_{i,j}=\left[ \bra{\psi_0}  U^{j-i} \ket{\psi_0}   \right]_{i,j},
}
since $(U^{i})^\dagger$ is the unique inverse of $U^i$, and therefore $(U^{i})^\dagger = U^{N-i}=U^{-i}.$
A matrix is circulant if $A_{i,j}=A_{i+k,j+k}$ where the addition is taken modulo $N$. The Gram matrix of the symmetric states is circulant, since it depends only on the difference $(j-i)$.

We should also note that we can write $U$ as
\EQ{U=\sum_{k=0}^{D-1}\exp(2\pi Ik/N)\ket{\gamma_k}\bra{\gamma_k},}
where $\{|\gamma_k\rangle\}_D$ is an orthonormal basis and $D$ is the dimension of the space spanned by the $\ket{\psi_i}$. We therefore have $\bra{\gamma_k}\gamma_{k'}\rangle=\delta_{k,k'}$. Note, that in general $N\neq D$, and it is important to keep track of in what range each index is defined. For the special case of linearly independent symmetric states, $N=D$ and the derivations simplify. By expressing $\ket{\psi_0}$ in terms of $\ket{\gamma_k}$,
\EQ{\ket{\psi_0}=\sum_{k=0}^{D-1} b_k\ket{\gamma_k},}
we obtain
\EQ{\ket{\psi_i}=\sum_{k=0}^{D-1}b_k\exp(2\pi I i k/N)\ket{\gamma_k}.}
We can then express the Gram matrix $G$ which is $N\times N$ matrix, in terms of a matrix $M$ which is $D\times N$ matrix,
\EQ{G = M^\dagger M,}
where
\EQ{ M=\left( \begin{array}{cccc}
\bra{\gamma_0}\psi_0\rangle, &\bra{\gamma_0}\psi_1\rangle, & \cdots, & \bra{\gamma_0}\psi_{N-1}\rangle\\
\bra{\gamma_1}\psi_0\rangle, & \bra{\gamma_1}\psi_1\rangle, & \cdots, & \bra{\gamma_1}\psi_{N-1}\rangle\\
\cdots, & \cdots, & \cdots, & \cdots\\
\bra{\gamma_{D-1}}\psi_0\rangle, & \bra{\gamma_{D-1}}\psi_1\rangle, & \cdots, & \bra{\gamma_{D-1}}\psi_{D-1}\rangle\end{array} \right).}
The columns of $M$ are representations of the $\ket{\psi_i}$'s in the $\ket{\gamma_k}$ basis. We have 
\EQ{[M]_{i,j}=\bra{\gamma_i}\psi_j\rangle=b_i\exp(2\pi I ij/N).}
The Gram matrix, being circulant, can be diagonalised with the unitary discrete fourier transform $F$,
\EQ{F_{i,j}= 1/\sqrt{N} \exp (-2 \pi I i j/N ),}
and therefore
\EQ{F^\dagger G F = F^\dagger M^\dagger M F = (M F)^\dagger M F  = \Lambda,}
where $\Lambda$ is a diagonal matrix with the eigenvalues $\lambda_k$ of $G$ on the diagonal. With the above definitions we can see that
\EQ{[MF]_{i,k}&=&\sum_j[M]_{i,j}[F]_{j,k}=\sum_j b_i\exp(2\pi I ij/N)1/\sqrt N\exp(-2\pi I jk/N)\nonumber\\ 
&=& b_i\delta_{i,k}\sqrt N,}
which leads to
\EQ{\label{e-values}\lambda_i&=&N|b_i|^2\textrm{ for } i<D\nonumber\\
\lambda_i&=& 0\textrm{ otherwise}.}
In the derivation above we used the fact that  $\sum_j [ \exp (2 \pi I (i-k)j /N)]=N\delta_{i,k}$.
We can now rewrite the initial states $\ket{\psi_i}$ in terms of the eigenvalues of the Gram matrix,
\EQ{\ket{\psi_i}=1/\sqrt N\sum_{k=0}^{D-1}\sqrt{\lambda_k}\exp(2\pi I ik/N)\ket{\gamma_k}.}
In the basis of the $\ket{\gamma_k}$, the average operator $\Phi$ in Eq. (\ref{eq:phi}) becomes
\begin{eqnarray}
\Phi &=& 1/N \sum_{k=0}^{N-1} \sum_{i,j=0}^{D-1} \sqrt{\lambda_i \lambda_j}   \exp (2 \pi I i k /N) \exp (-2 \pi I j k /N) \ket{\gamma_i}\bra{\gamma_j} \nonumber\\
&=& 1/N \sum_{k=0}^{N-1} \sum_{i,j=0}^{D-1} \sqrt{\lambda_i \lambda_j}   \exp (2 \pi I (i-j)k /N)  \ket{\gamma_i}\bra{\gamma_j}\nonumber\\&=&\sum_{i=0}^{D-1} \lambda_i    \ket{\gamma_i}\bra{\gamma_i},
\end{eqnarray}
 where we in the last step used the fact that $\lambda_i$ are all non-negative.
In this basis, the average operator is  thus diagonal and the elements on the diagonal are the first $D$ eigenvalues of the Gram matrix. Since the first $D$ eigenvalue are non-zero (are related with the $D$-coefficients $b_i$ from eq. (\ref{e-values}), which can be taken to be non-zero), the inverse in this basis is diagonal with elements $1/\lambda_i$. Therefore Eq. (\ref{SRM-states}) becomes
\EQ{\ket{\phi_i}=\Phi^{-1/2}\ket{\psi_i}=1/\sqrt N \sum_{k=0}^{D-1}\exp(2\pi I ik/N)\ket{\gamma_k},}
which are the DFT transformed $\ket{\gamma_k}$'s. We now obtain 
  \EQ{
   B_{i,j}=|\bra{\psi_i}\phi_j\rangle|^2=(1/N^2)| \sum_{k=0}^{D-1} \sqrt{\lambda_k} \exp (2 \pi  I  (j-i)k /N) |^2.}
It is worth mentioning that the operator $B$ with the matrix elements  $B_{i,j}$
is both circulant and symmetric.

The cost of making the SRM, for a cost matrix $C_{i,j}$, is given by
\EQ{\bar C_{SRM}=\sum_{i,j=0}^{N-1} \eta_i B_{i,j}C_{i,j}.}
We will see later that under certain circumstances this is also the minimum cost. For now, let us assume that the prior probabilities are equal, $\eta_i=1/N$, and that the cost matrix is circulant, i.e. that the matrix elements obey $C_{i,i+k}=C_{j,j+k}=\sum_k c_k \delta_{k,j-i}$. We then obtain
\EQ{\label{symmetric_min_cost_circulant}\bar C_{SRM}=1/N^2\sum_{k=0}^{N-1} c_k |\sum_{l=0}^{D-1}\sqrt{\lambda_l}\exp(2\pi Ikl/N)|^2.}
The minimum-error probability, which is the cost for $C_{i,j}=1-\delta_{i,j}$, i.e. $c_k=1-\delta_{k,0}$, becomes
\EQ{\label{min-error-sym} p_{min}= 1- (1/N^2) | \sum_{i=0}^{D-1} \sqrt{\lambda_i} |^2.}

\subsection{When is the SRM the minimum-cost measurement?}

In this section we will investigate under what conditions the minimum-cost measurement for $N$ symmetric states is the SRM, with a minimum cost given by Eq. (\ref{symmetric_min_cost_circulant}). In particular, we will examine the Helstrom conditions separately, and see what sufficient conditions we can impose on the cost matrix, such that the SRM is the optimal minimum-cost measurement.
For circulant and symmetric cost matrices, the three first Helstrom conditions are satisfied by the SRM, as shown in supplementary material of~\cite{Expr}. Here we will give an easier way to prove those conditions. We will then show that if the cost matrix obeys one more condition, then the last Helstrom condition, the inequality, also holds for the SRM, and thus the minimum-cost measurement for this type of cost matrices is the SRM.

\TH\label{1-3-conditions-circulant}
Let $\rho_i=\ket{\psi_i}\bra{\psi_i}$ be $N$ 
symmetric pure states, with equal prior probabilities $\eta_i=1/N$, and let $C_{i,j}$ be an $N \times N$ cost matrix which is circulant and symmetric. 
The three first Helstrom conditions for minimum-cost measurements can be re-written as the three first Helstrom conditions for a minimum-error measurement of the modified states $\rho'_i:=\sum_j C_{i,j}\rho_j$. That is,
\EQ{\label{helstrom_1}\Pi_i(\rho'_i-\rho'_j)\Pi_j=0}
for all $i,j$. This condition holds for $\Pi_i=\ket{\phi_i}\bra{\phi_i}$ ,which is the SRM for the initially considered pure states.
\HT
\proof
Eq. (\ref{helstrom_1}) becomes
\EQ{\sum_{k=0}^{N-1} C_{i,k}\bra{\phi_i}\psi_k\rangle\bra{\psi_k}\phi_j\rangle-\sum_{l=0}^{N-1} C_{j,l}\bra{\phi_i}\psi_l\rangle\bra{\psi_l}\phi_j\rangle=0}
We can find, for every term in the first sum, a corresponding term in the second sum, so that these terms cancel. 
The elements of the cost matrix $C_{i,j}$ and the terms $\bra{\phi_i}\psi_j\rangle$ depend only on the difference $j-i$ of the two indices, and it also holds that
\EQ{\bra{\phi_i}\psi_j\rangle=\bra{\psi_i}\Phi^{-1/2}\ket{\psi_j}=\bra{\psi_i}\phi_j\rangle=\bra{\psi_{i+l}}\phi_{j+l}\rangle.}
Therefore, each term with a given $k$ in the first sum, will be exactly cancelled by the term with $l=i+j-k$ in the second sum (recall that addition in indices is modulo $N$). Therefore the whole sum vanishes. 
We can see this by first noting that
\EQ{\bra{\phi_i}\psi_l\rangle\bra{\psi_l}\phi_j\rangle&=&\bra{\phi_i}\psi_{i+j-k}\rangle\bra{\psi_{i+j-k}}\phi_j\rangle=\nonumber\\
=\bra{\phi_k}\psi_j\rangle\bra{\psi_i}\phi_k\rangle&=&\bra{\psi_k}\phi_j\rangle\bra{\phi_i}\psi_k\rangle.}
What remains 
is to show that $C_{i,k}=C_{j,l}$ for $l=i+j-k$. This is the case because by assumption the cost matrix is both circulant and symmetric,
\EQ{C_{j,l}=C_{j,i+j-k}=C_{k,i}=C_{i,k}.}\qed
\\We now proceed to 
investigate when the fourth Helstrom condition holds.

\TH\label{thm-inequality-circulant}
Consider a collection of $N$ equiprobable symmetric states $\ket{\psi_i}$. If the cost matrix $C$ is (1) symmetric, $C_{i,j}=C_{j,i}$, (2) circulant, $C_{i,i+k}=C_{j,j+k}=c_k$, (3) the coefficients $c_k$ are non-positive, $c_k\leq0\quad\forall\quad k$ and (4) the cost matrix is negative semidefinite (its eigenvalues are all non-positive), then the SRM satisfies the inequality Helstrom condition for the minimum-cost measurement for the above cost matrix. Therefore, since the first three conditions are satisfied by theorem \ref{1-3-conditions-circulant}, the SRM is the minimum-cost measurement.
\HT
\proof
First, note that the eigenvalues of a circulant matrix are given by the discrete Fourier transform of the coefficients $c_k$. Thus the fourth condition of the above theorem reads
\EQ{\bar c_n=\sum_{k=0}^{N-1} c_k\exp(2\pi I k n/N)\leq0\quad\forall\quad n.}
The Helstrom inequality condition is
\EQ{\sum_{k=0}^{N-1} \eta_j C_{j,k}\rho_k-\sum_{i,k=0}^{N-1}\eta_i\Pi_i C_{i,k}\rho_k\geq0,}
where $\eta_i=1/N$. To prove that the operator in the LHS is positive definite, we need to prove that if we ``sandwich'' it 
with any general state $\ket{\chi}$, this always gives a positive number. We write 
\EQ{\ket{\chi}=\sum_{k=0}^{D-1} a_k\ket{\gamma_k},}
where $\ket{\gamma_k}$ is the $D$-dimensional orthonormal basis that we used earlier, that is, the Fourier transform of the basis $\ket{\phi_i}$ of the SRM. The Helstrom inequality condition becomes
\EQ{\label{inequality_circulant}\sum_{j=0}^{N-1}\sum_{k_1,k_2=0}^{D-1}C_{i,j}a^*_{k_1}a_{k_2}\bra{\gamma_{k_1}}\psi_j\rangle\bra{\psi_j}\gamma_{k_2}\rangle
-\nonumber\\-\sum_{m,j}^{N-1}\sum_{k_1,k_2}^{D-1}C_{m,j}a^*_{k_1}a_{k_2}\bra{\gamma_{k_1}}\phi_m\rangle\bra{\phi_m}\psi_j\rangle\bra{\psi_j}\gamma_{k_2}\rangle\geq0.}
We use the same definitions of $\Pi_i,\ket{\phi_i},\ket{\psi_i}$ as in the previous section. Moreover, note that since the cost matrix is circulant and symmetric, we have
\EQ{c_k=C_{i,i+k}=C_{i+k,i}=C_{i,i-k}=c_{-k}.}
We call the first term of eq. (\ref{inequality_circulant}) $A$ and the second term $B$. By using the definitions we obtain
\EQ{A&=&1/N\sum_{j=0}^{N-1}\sum_{k_1,k_2=0}^{D-1}C_{i,j}a^*_{k_1}a_{k_2}\sqrt{\lambda_{k1}\lambda_{k_2}}\exp(2\pi I (k_1-k_2)j/N)\nonumber\\
&=&1/N\sum_{l=0}^{N-1}\sum_{k_1,k_2=0}^{D-1}c_la^*_{k_1}a_{k_2}\sqrt{\lambda_{k1}\lambda_{k_2}}\exp(2\pi I (k_1-k_2)(l+i)/N),}
where on the second line, we have used $l=j-i$ and $C_{i,i+l}=c_l$. 
We also obtain
\EQ{B&=&1/N^2\sum_{m,j=0}^{N-1}\sum_{k_1,k_2=0}^{D-1}C_{m,j}a^*_{k_1}a_{k_2}\exp(2\pi Ik_1m/N)\times\nonumber\\ & &\times\left(\sum_{k_3=0}^{D-1}\sqrt{\lambda_{k_3}}\exp(2\pi Ik_3(j-m)/N)\right)\sqrt{\lambda_{k_2}}\exp(-2\pi Ik_2j/N)\nonumber\\
&=&1/N^2\sum_{k_1,k_2,k_3=0}^{D-1}a^*_{k_1}a_{k_2}\sqrt{\lambda_{k_2}\lambda_{k_3}}\times\nonumber\\ & &\times\sum_{m,j=0}^{N-1}C_{m,j}\exp(2\pi I m(k_1-k_3)/N)\exp(2\pi Ij(k_3-k_2)/N).}
Writing $C_{m,j}=c_l$, where $l=m+l$, and using the fact that
${\sum_m \exp(2\pi I m(k_1-k_2) /N)=N\delta_{k_1,k_2}}$,
we obtain
\EQ{B&=&1/N^2\sum_{k_1,k_2,k_3=0}^{D-1}a^*_{k_1}a_{k_2}\sqrt{\lambda_{k_2}\lambda_{k_3}}\times\nonumber\\
& &\times\sum_{m,l=0}^{N-1}c_l\exp(2\pi I m(k_1-k_3)/N)\exp(2\pi I(m+l)(k_3-k_2)/N)\\
&=&1/N\sum_{k_1,k_3=0}^{D-1}|a_{k_1}|^2\sqrt{\lambda_{k_1}\lambda_{k_3}}\left[\sum_{l=0}^{N-1} c_l\exp(2\pi Il(k_3-k_1)/N)\right].}
We now take $A-B$, renaming $k_3$ as $k_2$,
\EQ{A-B&=&1/N\sum_{k_1,k_2=0}^{D-1}\sqrt{\lambda_{k_1}\lambda_{k_2}}\left[\sum_{l=0}^{N-1} c_l\exp(2\pi Il(k_1-k_2)/N)\right]\times\nonumber\\& &\times\left[a^*_{k_1}a_{k_2}\exp(2\pi Ii(k_1-k_2)/N)-|a_{k_1}|^2\right].}
The above expressions followed since $C_{i,j}$ is symmetric, which implies that
\EQ{\bar c_n=\sum_{l=0}^{N-1} c_l\exp(2\pi Iln/N)=\bar c_{-n}.}
The fourth condition of the theorem states that $\bar c_n$, the eigenvalues of the cost matrix, are always negative. Therefore Eq. (\ref{inequality_circulant}) can further be written as (note that the remaining sums, in the following equations, take values from $k=0$ to $k=D-1$)
\begin{eqnarray} & &\sum_{k_1,k_2}|\bar c_{k_1-k_2}|\sqrt{\lambda_{k_1}\lambda_{k_2}}\left(|a_{k_1}|^2-a^*_{k_1}a_{k_2}\exp(2\pi Ii(k_1-k_2)/N)\right)\nonumber\\
& &
=\frac12\sum_{k_1,k_2}|\bar c_{k_1-k_2}|\sqrt{\lambda_{k_1}\lambda_{k_2}}
\times\nonumber\\ & &\times\left(|a_{k_1}|^2+|a_{k_2}|^2-a^*_{k_1}a_{k_2}\exp(2\pi Ii(k_1-k_2)/N)-a_{k_1}a^*_{k_2}\exp(-2\pi Ii(k_1-k_2)/N)\right)\nonumber\\& &
=\frac12\sum_{k_1,k_2}|\bar c_{k_1-k_2}|\sqrt{\lambda_{k_1}\lambda_{k_2}}
\left(|a_{k_1}|^2+|a_{k_2}|^2-2Re[a^*_{k_1}a_{k_2}\exp(2\pi Ii(k_1-k_2)/N)]\right)\nonumber\\
& &\geq\frac12\sum_{k_1,k_2}|\bar c_{k_1-k_2}|\sqrt{\lambda_{k_1}\lambda_{k_2}}
\left(|a_{k_1}|^2+|a_{k_2}|^2-2|a_{k_1}||a_{k_2}|\right)\nonumber\\
& &=\frac12\sum_{k_1,k_2}|\bar c_{k_1-k_2}|\sqrt{\lambda_{k_1}\lambda_{k_2}}(|a_{k_1}|-|a_{k_2}|)^2\geq0
\end{eqnarray}
which completes the proof. Note that (a) we have multiplied the expressions with $N$, (b) in the second line we used the general property $\sum_{k_1,k_2}L_{k_1,k_2}=1/2\sum_{k_1,k_2}(L_{k_1,k_2}+L_{k_2,k_1})$, where $L$ was the full expression 
in the sum over $k_1,k_2$, and (c) the inequality from the third to te forth line comes from the property $Re[z_1z_2]\leq|z_1||z_2|$ of complex numbers.\qed

To illustrate what conditions on $c_k$'s are imposed by the requirement that the eigenvalues of the cost matrix are all non-positive, we consider the case $N=4$:
\begin{eqnarray}\bar c_0&=&c_0+c_2+2c_1\nonumber\\
\bar c_1&=&c_0-c_2\nonumber\\
\bar c_2&=&c_0+c_2-2c_1\nonumber\\
\bar c_3&=&c_0-c_2=\bar c_1,
\end{eqnarray}
where we have used that $c_1=c_3$. Given that $c_0,c_1,c_2,c_3\leq0$, the SRM will be the  minimum-cost measurement for this cost matrix, if
\EQ{c_2\geq c_0\textrm{ and }c_1\geq \frac{c_0+c_2}2.}


\subsection{Example: Bounding the minimum cost using SRM for coherent symmetric states} 

Here we will consider an example of four symmetric coherent states, given by $\{\ket{\alpha},\ket{i\alpha},\ket{-\alpha},\ket{-i\alpha}\}$, for amplitude $\alpha=2$. This symmetric set of states occurs in an implementation of quantum digital signatures~\cite{QDS_Exp}. The choice of protocol parameters, such as signature length, in order to guarantee sufficient security, depends on the ability of a malevolent party to forge a message. This in turn depends on the minimum cost of the best measurement a malevolent party could make on all signature copies they can obtain. In finding a bound for how well signed messages can be forged, it is crucial to bound the minimum cost for a generic cost matrix (which in general comes from  experimental parameters). We will give a method for how to obtain such bounds, using, as an example, a cost matrix that was actually obtained in an experiment on quantum digital signatures \footnote{The actual data that was used in that work was slightly different. The technique used to bound the forging probability was similar, but not identical, to the one presented here. We chose to use this data  to better illustrate the use of the results presented in this paper.} \cite{QDS_Exp}. 
This cost matrix is given by
\EQ{\label{cost-exp}C=\left( \begin{array}{cccc}
9.34\times10^{-5}, & 7.81\times10^{-4}, & 1.19\times10^{-3}, & 8.70\times10^{-4}\\
9.53\times10^{-4}, & 3.25\times10^{-4}, & 9.74\times10^{-4}, & 1.36\times10^{-3}\\
1.43\times10^{-3}, & 1.40\times10^{-3}, & 6.35\times10^{-5}, & 9.61\times10^{-4}\\
8.10\times10^{-4}, & 1.62\times10^{-3}, & 9.38\times10^{-4}, & 7.07\times10^{-5}\end{array} \right).}
One can of course numerically compute the minimum cost using semi-definite programming. However, here we provide some analytical bounds using the properties we derived above, and the expressions for the SRM.

Before attempting to bound the minimum cost, we will first compute the SRM states $\ket{\phi_i}$ for this case and the corresponding minimum-error probability. The elements of the Gram matrix are given by
\EQ{\bra{\alpha}\alpha\rangle=1&,&~~\bra{\alpha}i\alpha\rangle=\exp(-\alpha^2(1-i)),\nonumber\\
\bra{\alpha}-\alpha\rangle=\exp(-2\alpha^2)&,&~~\bra{\alpha}-i\alpha\rangle=\exp(-\alpha^2(1+i)).}
Its eigenvalues are calculated as
\EQ{\lambda_1&=&2\exp(-\alpha^2)(\cos(\alpha^2)+\cosh(\alpha^2))\\
\lambda_2&=&2\exp(-\alpha^2)(\sin(\alpha^2)+\sinh(\alpha^2))\\
\lambda_3&=&2\exp(-\alpha^2)(\cosh(\alpha^2)-\cos(\alpha^2))\\
\lambda_4&=&2\exp(-\alpha^2)(\sinh(\alpha^2)-\sin(\alpha^2)).}
From this we can now write the states $\ket{\phi}$ using the Fourier orthonormal basis $\ket{\gamma_k}$,
\EQ{\ket{\phi_j}=\frac1{\sqrt{N}}\sum_i\exp(2\pi I ij/N)\ket{\gamma_i},}
and the $B_{i,j}$ as
\EQ{B_{i,j}=\frac1{16}|\sum_l\sqrt{\lambda_l}\exp(2\pi I(j-i)l/4)|^2.}
The minimum error is then given by
\EQ{p_{min}=1-1/16|\sum_i\sqrt{\lambda_i}|^2=0.000168.}
We now return to the minimum-cost measurement for the cost matrix in Eq. (\ref{cost-exp}). In order to analytically bound the minimum cost using the methods given in the previous sections, we follow five steps.

\begin{enumerate}

\item We rewrite the cost matrix $C$ as sum of a constant-row matrix $C^h$ and the smallest possible non-negative remaining matrix $C'$. This is achieved by subtracting, from all elements of each row, the smallest element on that row. The cost for the constant-row matrix $C^h$ is the smallest cost one can  possibly obtain, even if one knows what state is actually sent, and  
is given by 
$\bar C^h\sum_i\eta_i\min_j C_{i,j}$. For our example, the smallest cost in every row is on the diagonal. Thus the cost for $C^h$ is $\bar C^h=1/4\sum_i C_{i,i}=1.38\times10^{-4}$. We obtain the matrix
\EQ{ C'=\left( \begin{array}{cccc}
0, & 6.88\times10^{-4}, & 1.10\times10^{-3}, & 7.77\times10^{-4}\\
6.28\times10^{-4}, & 0, & 6.49\times10^{-4}, & 1.04\times10^{-3}\\
1.37\times10^{-3}, & 1.34\times10^{-3}, & 0, & 8.98\times10^{-4}\\
7.39\times10^{-4}, & 1.55\times10^{-3}, & 8.68\times10^{-4}, & 0\end{array} \right).}

\item We further subtract the greatest fully constant matrix with $C^c_{i,j}=M$ for all $i,j$, so that the remaining cost matrix is strictly non-positive, i.e. $C'_{i,j}=M+C''_{i,j}$. This means subtracting, from all elements $C'_{i,j}$, the greatest element in that matrix.  For our example, the greatest element is $1.55\times10^{-3}=M$, and this leads to (note the minus sign outside the matrix)
\EQ{ C''=-\left( \begin{array}{cccc}
1.55\times10^{-3}, & 0.86\times10^{-3}, & 0.45\times10^{-3}, & 0.77\times10^{-3}\\
0.92\times10^{-3}, & 1.55\times10^{-3}, & 0.90\times10^{-3}, & 0.51\times10^{-3}\\
0.18\times10^{-3}, & 0.21\times10^{-3}, & 1.55\times10^{-3}, & 0.65\times10^{-3}\\
0.81\times10^{-3}, & 0, & 0.68\times10^{-3}, & 1.55\times10^{-3}\end{array} \right).}
The overall cost so far is $\bar C(\Pi)=\bar C^h+M+\bar C''(\Pi)$, where the cost of $C''$ is a function of the measurement made, and takes a negative value, since all the elements of the matrix are negative. 

\item The cost of any cost matrix which is smaller, element by element, than $C''$, bounds the overall cost from below. To find the tightest bound, we look for such a matrix with the largest possible elements, which also satisfies the conditions of theorem \ref{thm-inequality-circulant}, so that the minimum cost is given by the SRM. For our example, the largest cost matrix which is smaller than $C''$ and is circulant, symmetric and has negative eigenvalues, is given by $C^l=\{c_0=-1.55\times10^{-3},c_1=-0.92\times10^{-3},c_2=-0.51\times10^{-3}\}$. Note that the condition for negative eigenvalues is satisfied, $c_2\geq c_0$ and $c_1\geq\frac{c_0+c_2}2$. It follows that the SRM gives the minimum cost for $C^l$, and this cost is $\bar C^l_{min}=-1.54989\times10^{-3}$. This gives a lower bound for the minimum cost of $C$,    \EQ{\bar C_{min}\geq \bar C^h+M+\bar C^l_{min}=1.38\times10^{-4}+1.1\times10^{-7}.}

\item Similarly, to find an upper bound, we seek a cost matrix which is larger than $C''$, element by element, which is the smallest possible matrix which also satisfies the conditions of theorem \ref{thm-inequality-circulant}. This matrix is given by $C^u=\{c_0=-1.55\times10^{-3},c_1=-0.21\times10^{-3},c_2=0\}$. We can also confirm that its eigenvalues are negative, since the conditions for this are satisfied. Therefore the SRM is the minimum-cost measurement for $C^u$, with the cost $\bar C^u_{min}=-1.54978\times10^{-3}$. This leads to an upper bound for the minimum cost of $C$ as
    \EQ{\bar C_{min}\leq \bar C^h+M+\bar C^u_{min}=1.38\times10^{-4}+2.2\times10^{-7}.}
    \end{enumerate}
We therefore obtain the bounds
\EQ{1.38\times10^{-4}+2.2\times10^{-7}\geq \bar C_{min}\geq 1.38\times10^{-4}+1.1\times10^{-7}.}
We see that these bounds are relatively tight. The minimum cost is of the order of $10^{-4}$, while the accuracy that the minimum cost is bounded by is of order $10^{-7}$. Another point to mention is that in the case the cost matrix after subtracting the constant-row $C^h$ is circulant, then it is likely that the two bounds coincide. In other words, in that case we obtain the exact minimum cost. 
A final point to stress here is that both the upper and lower bounds are important for different type of circumstances. If, for example, the minimum cost corresponds to the probability that some malevolent party correctly guesses the state, thereby undermining the security of some cryptographic protocol, then we are interested in the worst-case scenario, which is that he makes the best possible guess. We then use the lower bound of the minimum cost in order to make sure that our protocol is secure. If, on the other hand, some honest party  is required to make the guess, then the worst case scenario corresponds to the upper bound for the minimum cost.

\section{Minimum-error measurement and probabilities for mixed states of symmetric pure states}

In the previous section we have seen that the minimum-cost measurement for a wide class of cost matrices for symmetric pure states is the SRM. More specifically, using the results of section \ref{Section-general-results-cost}, we see that if we can make a cost matrix circulant, with non-positive entries and negative semidefinite, by adding (subtracting) constant-row matrices, then the minimum-cost measurement is the SRM. Moreover, the cost can be easily analytically computed using the expressions for the SRM in terms of the eigenvalues of the Gram matrix of the symmetric states.

Here we will use the above result, and the equivalence between  minimum-cost measurements for  pure states and  minimum-error measurements for mixed states, which we discussed in section \ref{Section-error-cost-general}, to obtain the minimum-error probability for a class of mixed states which are mixtures of pure symmetric states. We will similarly provide bounds on the minimum-error probability for a larger class of mixed states.

The first observation is that for any collection of mixed states of the form $\bar\rho_i=\sum_j a_{i,j}\ket{\psi_j}\bra{\psi_j}$, where $\ket{\psi_j}$ are symmetric states, we can rephrase any constraints (for a given measurement to be optimal) on the cost matrix with elements $C_{i,j}$ in terms of conditions on the $a_{i,j}$. In particular, assuming for simplicity that the prior probabilities $\eta_i$ of the different mixed states $\bar\rho_i$ are all equal to $1/N$, we obtain
\EQ{a_{i,j}=1-C_{j,i}.}
Requiring that the cost matrix $C$ is symmetric and circulant implies that the matrix with elements $a_{i,j}$ should also be symmetric and circulant, while requiring that the cost matrix $C$ is negative semidefinite, implies the requirement that $a_{i,j}$ define a positive semidefinite matrix. 
The results in the previous section imply that if the states $\bar\rho_i$ are such that the $a_{i,j}$ define a circulant, symmetric and positive definite matrix, then the SRM is the minimum-error measurement for the mixed states $\bar\rho_i$'s.

An interesting thing to point out is that mixed states generated by a circulant, symmetric matrix $a_{i,j}$, from pure symmetric states, are also symmetric states, induced by the same symmetry unitary. We can see that, since
\EQ{U\bar\rho_iU^\dagger&=&\sum_j a_{i,j}U^\dagger\ket{\psi_j}\bra{\psi_j}U^\dagger=\sum_ja_{i,j}\ket{\psi_{j+1}}\bra{\psi_{j+1}}\\
&=&\sum_j a_{i+1,j+1}\ket{\psi_{j+1}}\bra{\psi_{j+1}}=\sum_ja_{i+1,j}\ket{\psi_j}\bra{\psi_j}=\bar\rho_{i+1}.}
We have therefore shown that the SRM  is the minimum-error measurement, even for mixed symmetric states defined as above, provided the eigenvalues of $a_{i,j}$ are non-negative. 
This is in agreement with the result of ref. \cite{Chou_Hsu}.

An other interesting consequence concerns the case where the mixed states are arbitrary mixtures of symmetric states. In other words, when the matrix defined by $a_{i,j}$ is more general. As we have outlined in the example in the previous section, we are able to provide upper and lower bounds for the minimum error of those mixed states (given by the minimum cost for the corresponding pure states), using the explicit and easy form of the SRM for symmetric pure states. In particular, if those bounds are accurate, 
compared to other significant parameters that may interest us, then we can use the bounds provided by the SRM to estimate the minimum-error probability for the mixed states.


Finally, one should note that if the analytical form of the minimum-cost measurement for some class of cost matrices is known (as in our examples the SRM), then one can obtain bounds for the minimum error for a related class of mixed states using the methods we described.


\section{\label{Section-tensor-cost}Minimum cost for sequences of states}

In this section we will consider tensor products of states. In particular, we will focus on a special case, which is important for quantum cryptography. The Hilbert space is a tensor product of identical Hilbert spaces $\hil_{tot}=\otimes_{i=1}^L \hil_i$.
We refer to the whole state as global, and the individual states as local.  The set of possible states that we are going to consider consists of all (tensor product) combinations of the $N$ different local states, for the $L$ different subsystems that make the global state.

Such 
states occur frequently in quantum information science. The local states comprise an alphabet of possible quantum ``letter" states, 
whereas the total tensor product state form a quantum ``message". Such states occur in, for example, QKD, where the total system Alice sends to Bob is a sequence of $L$ local states. In BB84, the local states belong to two mutually unbiased bases. Appropriately ordered, the states form a set of symmetric states. Analogous situations occur in quantum digital signatures, universal blind quantum computing, and other protocols. Considering the entire global system, as opposed to individual components, which was the topic of previous sections, leads to collective (or coherent) measurement strategies which can be uses to gain information about the system.

To each individual local system one can assign a local minimum-cost problem, which is the situation we discussed previously. From the collection of local problems, one can derive a global minimum-cost problem, where the global cost is some function of local costs. A typical example of this is the scenario in which a party wishes to identify the message sent, in a way which minimizes the number of local states for which a misidentification occurred.
In this paper we will consider the more general case of global cost matrices where the cost for each global state is some (general) function of the sum of the (local) costs of the subsystems. We further assume that the local cost matrices are all identical for the different subsystems. This type of systems and cost matrices are widely used. 

The question of whether the optimal measurement is a tensor product of local measurements, in scenarios where the possible states are tensor product states, was crucial in the development of QKD. The optimal measurement 
for obtaining the parity of a bit string, in the context of QKD, was examined by Fuchs and Graaf in \cite{FG} and by Bennett, Mor and Smolin in \cite{BMS}. 
It turns out that whether or not a sequence of local measurements is optimal depends on the global cost matrix, that is, on the specific global cost function. 
In particular, it was shown that the parity of a string of bits, encoded in qubits as in the BB84 protocol, can be best guessed by measuring in an entangled basis. The parity of a string is equal to addition modulo 2 of the bits, and the global cost becomes a function of the local costs.

It may seem counter-intuitive that entangled measurements outperform local ones, since the possible states are all tensor products, and there are no correlations between individual bits or qubits in the example with the parity. 
However, for correctly determining the parity of the string of bits in this example, there is no optimum ``local'' measurement strategy. If we obtain the correct bit value after measuring the first qubit, then the best strategy is to guess the second bit correctly. But if we have guessed the first bit wrong, it is beneficial to make another mistake for the second one so that the parity is guessed correctly. The overall cost is a periodic function of the sum of the local costs. What is more surprising, however, is that even if the global cost is a monotone function of the local costs, 
then it is still not guaranteed that the global optimal measurement is non-entangled.

In this section we will first prove that for a total cost matrix which is a linear function of the sum of the local costs, the minimum-cost measurement is a tensor product of local measurements. We will then provide bounds for total costs which are convex and concave functions of the sum of local costs. Finally we will give an example of a monotone function, a step function, for which the minimum-cost measurement is a measurement in an entangled basis. This example is interesting for various reasons. First, this type of cost matrix appears in protocols for QDS. Second, it is closely related to conclusive state elimination~\cite{stevebook,OppenUSE}. Third, this type of measurement is the one used to argue that an epistemic view of the wavefunction is impossible \cite{PBR}. 

We should introduce some notation here. The total number of local subsystems is $L$, and we call the global space of all subsystems $\Omega$. We label the global possible states as $\rho^{tot}_k=\otimes_i \rho_{k(i)}$. We will use the index $k$ for the global space, that is, it takes $N^L$ different values. To refer to different such global states, we will use subscripts (e.g. $k_1,k_2,\cdots$). When we want to refer to the state of a particular subsystem, 
e.g. the $i$'th, we will write $k(i)$.
We assume that each subsystem is identical, and has $N$ different possible states. The states of the subsystems are independent of each other, so that 
the prior probabilities for the global states can be written as products $\eta^{tot}_k=\prod_i \eta_{k(i)}$. Note that $\sum_{k(i)}\eta_{k(i)}=1$ for all $i$, since the probabilities of each subsystem sum to one.

The cost matrices we are considering have entries of the form $C_{k_1,k_2}= f(\sum_i C^i_{k_1(i),k_2(i)})$. $C_{k_1, k_2}$ is the cost of choosing outcome $k_2$ if the global state was $\rho_{k_1}$. $C^i_{k_1(i),k_2(i)}$ are the entries of the local cost matrices. 
The cost of a global measurement corresponding to a POVM $\Pi$ with elements $\{\Pi_{k}\}$ is given by
\EQ{\label{tensor_min_cost}\bar C(\Pi)=\sum_{k_1,k_2}\eta_{k_1} C_{k_1,k_2}Tr(\Pi_{k_2}\rho_{k_1}).}
Indices in the above take values from one to  $N^L$, as they will always do, unless the particular element is specified. For example, $k_1(i)$ is the index  for the $i$'th subsystem, in the sequence $k_1$. The task here is to find under what conditions on $C_{k_1,k_2}$ (which is in our case is a function of the sum of the local costs) the minimum-cost measurement is 
to make optimal local minimum-cost measurements. For those cases, the value of the minimum cost can also be computed.


\subsection{Cost matrix in the form of a linear function of the sum of local costs}

\TH\label{linear_cost}
Assume a set of product states with independent prior probabilities for the subsystems. Assume that the global cost matrix $C_{k_1,k_2}$, is a linear function of the sum of some local cost matrices entries $C^i_{k_1(i),k_2(i)}$. In other words,

\EQ{C_{k_1,k_2}=f\left(\sum_i C^i_{k_1(i),k_2(i)}\right)=a\sum_iC^i_{k_1(i),k_2(i)}+b}
with $f(x)=ax+b$. Then (i) the minimum-cost measurement is the tensor product of the local minimum-cost measurements for the local costs $C^i$ and (ii) the minimum cost is given as $\bar C_{min}=a\sum_i \bar C^i_{min}+b$.
\HT
In order to prove the above theorem, we first need few lemmas.
\LE\label{indep_subspace_cost}
Consider a subset $A$ of $\Omega$ that consists of a collection of local subspaces $i\in A$ and call $\bar A=\Omega\setminus A$. The Hilbert space associated with $A$ is $\hil_A=\otimes_{i\in A}\hil_i$. Assume that the global cost matrix depends only on $i\in A$, i.e. $C_{k_1,k_2}=f(i\in A)$. Then for any global measurement $\Pi\in\hil_{tot}$, there exists another measurement of the form $\bar\Pi_{A}\otimes\one_{\bar A}$, with $\bar\Pi\in\hil_{A}$, that gives the same cost $\bar C(\Pi_\Omega)=\bar C(\bar\Pi_{A}\otimes\one_{\bar A})$.
\EL
\proof
First we should note that the prior probabilities are of the form $\eta_k=\eta_{k(A)}\eta_{k(\bar A')}$, i.e. independent for $\hil_A$ and $\hil_{\bar A}$. 
We will prove the lemma by explicit construction. From eq. (\ref{tensor_min_cost}) we obtain the following expression for the cost, where the subscripts 
for the POVMs 
indicate on which subsystems they act, 
\EQ{\bar C(\Pi_\Omega)&=&\sum_{k_1(A),k_1(\bar A)}\sum_{k_2(A),k_2(\bar A)}C_{k_1(A),k_2(A)}\eta_{k_1(A)}\eta_{k_1(\bar A)}\times\nonumber\\& &\times Tr\left(\Pi_{k_2(A),k_2(\bar A)}\rho_{k_1(A)}\otimes\rho_{k_1(\bar A)}\right).\label{eq:globalcost}}
An important thing to note is that the sums in Eq. (\ref{tensor_min_cost}) run over 
all $k_1,k_2$, where we have decomposed these sums to summing over the different possibilities for the subsystems (summing over $k_1(A),k_1(\bar A),k_2(A),k_2(\bar A)$). The operator $\Pi_{k_2}$ has also been expressed as function of $k_2(A)$ and $k_2(\bar A)$, without implying that it has product structure. Finally, note that the cost matrix, by the assumptions in the lemma, depends only on the indices belonging to $A$.

By defining a POVM which acts on $\hil_A$ (note the partial trace) as
\EQ{\bar\Pi_{k_2(A)}=Tr_{\bar A}\left(\Pi_{k_2}\cdot\one_A\otimes \left(\sum_{k_1(\bar A)}\eta_{k_1(\bar A)}\rho_{k_1(\bar A)}\right)\right),}
it follows that the lemma holds since one can easily check that
\EQ{\bar C(\Pi_\Omega)=\bar C(\bar\Pi_A\otimes\one_{\bar A}).}
Therefore, for all possible costs, one can find a measurement acting non-trivially only on $\hil_A$, achieving that cost. In the cases described by this lemma, with no loss of generality, for any optimization we can restrict out attention to measurements acting on $\hil_A$.\qed

\LE\label{product_cost_matrix}
If the cost matrix depends only on a subsystem $A$, i.e. $C_{k_1,k_2}=f(i\in A)$, and we have any measurement with a POVM of the form $\Pi_A\otimes\Pi_{\bar A}$, then the cost of the measurement is independent of the measurement on subsystem $\bar A$, that is,
\EQ{\bar C(\Pi_A\otimes\Pi_{\bar A})=\bar C(\Pi_A\otimes\Pi'_{\bar A})=\bar C(\Pi_A\otimes\one_{\bar A}).}
\EL
\proof Since both the state and the elements of the POVM, are factorizable, the trace is simply the product of the trace of the subsystems $A,\bar A$, and Eq. (\ref{tensor_min_cost}) becomes
\EQ{\bar C(\Pi_A\otimes\Pi_{\bar A})&=&\sum_{k_1(A),k_2(A)}\eta_{k_1(A)}C_{k_1(A),k_2(A)}Tr\left(\Pi_{k_2(A)}\rho_{k_1(A)}\right)\times\nonumber\\&&\times\left(Tr\left(\sum_{k_1(\bar A),k_2(\bar A)}\Pi_{k_2(\bar A)\rho_{k_1(\bar A)}}\right)\right)=\nonumber\\&=&\sum_{k_1(A),k_2(A)}\eta_{k_1(A)}C_{k_1(A),k_2(A)}Tr\left(\Pi_{k_2(A)}\rho_{k_1(A)}\right),}
where we have used the fact that $C_{k_1,k_2}$ is independent of $k_1(\bar A)$ and $k_2(\bar A)$ to move the second sum in (\ref{eq:globalcost}) inside the trace, and also that $\sum_{k_2(\bar A)}\Pi_{k_2(\bar A)}=\one_{\bar A}$, the trace of the density matrix is one and $\sum_{k_1(\bar A)}\eta_{k_1(\bar A)}=1$.\qed

\LE
If the global cost is constant function ($f(x)=C$) and therefore $C_{k_1,k_2}=C$, then all measurements give same cost equal to that constant $C$.\EL
This follows from the definition of cost.

\LE \label{lemma:product} Consider a global cost matrix that is equal to the sum of the local cost matrices $C_{k_1,k_2}=\sum_i C^i_{k_1(i),k_2(i)}$ (corresponds to the case of a function $f(x)=x$ of the sum of the individual cost matrices). Then, the minimum-cost measurement is given by tensor product of local minimum-cost measurements. Moreover the minimum cost is given by $\bar C_{min}=\sum_i \bar C^i_{min}$.\EL
\proof
Eq. (\ref{tensor_min_cost}) can be rewritten as
\EQ{C(\Pi)=\sum_{k_1,k_2}\eta_{k_1}(\sum_i C^i_{k_1(i),k_2(i)})Tr(\Pi_{k_2}\rho_{k_1})=\sum_i C^i(\Pi),}
where we defined
\EQ{C^i(\Pi)=\sum_{k_1,k_2}(\eta_{k_1}) C^i_{k_1(i),k_2(i)}Tr(\Pi_{k_2}\rho_{k_1}).}
Intuitively, each $C^i$ corresponds to a cost matrix that has no cost for any declaration for any subsystems except for subsystem $i$. The minimum cost of $C^i$ is denoted by $\bar C^i_{min}$. By noting that each $C^i$ depends only on the $i$'th element and using lemma \ref{indep_subspace_cost}, we have
\EQ{\bar C^i_{min}=\bar C^i(\Pi^{min}_i\otimes\one_{\Omega\setminus \{i\}})=\bar C^i(\Pi^{min}_i\otimes\Pi_{\Omega\setminus \{i\}}),}
where $\Pi_{\Omega\setminus\{i\}}$ is any element of a POVM acting on that space, and the second equality follows from Lemma \ref{product_cost_matrix}. Moreover, from Lemma \ref{sum of cost matrices}, it follows that the minimum total cost cannot be less than the sum of the minimum costs of each term in the sum. However, since for each term of the sum we have a measurement that has relevant support only on one subspace (the measurement on the remaining subsystems can be arbitrary), it is possible to have a measurement that achieves the minimum cost for all terms simultaneously, and thus the lower bound of lemma \ref{sum of cost matrices} can actually be achieved. The measurement is given by the operators $\otimes_i \Pi^{min}_i$, and gives the cost $\bar C_{min}=\sum\bar C^i_{min}$.\qed
\\Note that we have shown that there exists a minimum cost measurement that is local. Since the optimal measurement is not unique, there may also be non-local measurement that achieves the same minimum cost.

Finally, it follows that Theorem \ref{linear_cost} holds from the last two lemmas and the definition of the cost matrix.

\subsection{Convex, concave, monotonic and general functions}
Here we will consider bounds and statements which apply when the global cost matrix is a general function of the sum of some local costs.

\LE
Assume that we have a global cost matrix that is a convex function of the sum of some local costs $C_{k_1,k_2}=f(\sum_i C^i_{k_1(i),k_2(i)})$. Then the global minimum cost is upper bounded by the sum of local minimum costs, 
\EQ{C_{min}\leq\sum_i f(C^i_{min}).}
\EL
\proof
This follows by noting that $f((1/N)\sum_i C^i_{k_1(i),k_2(i)})\leq (1/N)\sum_i f(C^i_{k_1(i),k_2(i)})$, and by Lemma \ref{lemma:product}, which says that the minimum cost, for a global cost function which is a sum of local costs, is given by the sum of the local minimum costs for local cost functions 
$(1/N) f(C^i_{k_1(i),k_2(i)})$. The minimum cost obtained by making the local optimal measurements is therefore greater than or equal to the minimum possible cost for the cost function $C_{k_1,k_2}=f(\sum_i C^i_{k_1(i),k_2(i)})$, and thus provides an upper bound for the cost we are interested in. 
\qed
\LE
Assume that we have a global cost matrix which  is a concave function of the sum of some local costs, $C_{k_1,k_2}=f(\sum_i C^i_{k_1(i),k_2(i)})$. Then the global minimum cost is lower bounded by the sum of local minimum costs, 
\EQ{C_{min}\geq\sum_i f(C^i_{min}).}
\EL
\proof
This again follows by noting that $f((1/N)\sum_i C^i_{k_1(i),k_2(i)})\geq (1/N)\sum_i f(C^i_{k_1(i,k_2(i))})$, and by Lemma  \ref{lemma:product}. 
The minimum cost obtained by optimal local measurements for the local costs $(1/N)f(C^i_{k_1(i),k_2(i)})$ is less or equal to the minimum cost in question, and thus provide a lower bound for this minimum cost. \qed

\subsection{Functions for which local measurements are sub-optimal, state elimination, and the PBR argument}

In this section, we will give an example, which proves that even if the function of the local costs is monotonically increasing, the minimum cost measurement is not necessarily given by local minimum-cost measurements. We will consider a cost matrix which is a step function of the sum of the local costs. A step function is an important example, since in cryptographic protocols such as QDS \cite{QDS,Expr,QDS_VPE,QDS_Exp}, a party
will accept a signed message as genuine if it contains fewer mismatches than a particular threshold.
This means that achieving fewer mismatches than this threshold carries no cost, since the signed message is accepted as genuine, while  exceeding the threshold has cost equal to one, since the message is rejected.

Consider a sequence of two qubits, each of them is either in the state $\ket{0}$ or in the state $\ket{+}=1/\sqrt{2}(\ket{0}+\ket{1})$. The global cost is given by a step function of the sum of the local costs, where the local cost matrices are the error probability $C_{i,j}=1-\delta_{i,j}$. In particular, we will consider the case where if both bits are wrong then the cost is one, while if only one or none of the bits are wrong, then there is no cost at all. In other words, we want to be sure that we do not make mistake for both elements, but either zero or one error is fine.

The best local measurement is clearly to perform a minimum-error measurement for each qubit. The cost for this measurement is given by
\EQ{\bar C(local)=p^2_{min}=(1-1/2(\sqrt{1-|\bra{0}+\rangle|^2}+1))^2=0.021}
which is the minimum probability of error for both (independent) elements.
However, there exists a measurement in an entangled basis (we will call this the \emph{PBR basis}), that gives a smaller cost. If we measure in the following basis,
\EQ{\ket{\phi_{++}}=1/\sqrt2(\ket{01}+\ket{10})\\
\ket{\phi_{+0}}=1/\sqrt2(\ket{0-}+\ket{1+})\\
\ket{\phi_{0+}}=1/\sqrt2(\ket{+1}+\ket{-0})\\
\ket{\phi_{00}}=1/\sqrt2(\ket{+-}+\ket{-+}),}
then we will \emph{never} make two mistakes.
The cost for this measurement, $\bar C(PRB)$, is therefore exactly zero.

We should make two comments here. First, this measurement basis was given by Pusey, Barrett and Rudolph (PBR)~\cite{PBR} in an argument for proving that the nature of the wavefunction in quantum mechanics is not epistemic. Here we give a simplified version of this argument. The PBR argument started with the assumption that the wavefunction represents an epistemic distribution over some underlying different ontic states. Since the local states are non-orthogonal, they concluded that some ontic states are compatible with both $\ket{0}$ and $\ket{+}$ with some non-zero probability that is directly related to $p_{min}$. Having a pair of uncorrelated, non-interacting local states, would imply that there are some global ontic states, with probability $p_{min}^2$, that are compatible with all four possible wavefunctions $\{\ket{00},\ket{0+},\ket{+0},\ket{++}\}$. However, if one measures in the PBR basis, any outcome that is obtained is incompatible with (rules out) one of the four possible initial states. This manifest itself by the fact that the $\bar C(PBR)$ is zero. Therefore, the assumption that the wavefunction has purely epistemic character has to be rejected. It is very interesting that this deep philosophical insight is immediately connected to the security of cryptographic protocols.

The second comment is that this exact type of measurement can be understood as quantum state elimination or quantum state exclusion~\cite{stevebook,OppenUSE,QDS_Exp}. Depending on which of the four possible outcomes is obtained, we can, with 100\% probability, rule out one of the possible states. In particular, we can rule out the state for which both qubits are different compared with our result. This again is slightly counter-intuitive, since we started with four possible linearly independent non-orthogonal states. While it is well known that we cannot determine the state with certainty, we can rule out (eliminate) a state with certainty.

Finally, an interesting observation is that it is the inequality Helstrom condition that is expected to fail for the local measurements. In a sense, the local minimum-cost measurements corresonds to a ``local minimum'', in the sense that it is optimal compared to other slight perturbations. However, there is an entangled basis which is globally optimal. In the appendix \ref{app} we see that for sequences of symmetric states with a global cost matrix which is any function of the sum of the local costs, the three first conditions of Helstrom hold for local SRMs. It is the failure of inequality condition, however, that leads to an optimal measurement in an entangled basis for certain global cost functions\footnote{Note that even in the example with a step function, it is not always the case that global measurements outperform local ones. This depends on the particular value at which the step occurs. In the example 
we presented, if the step function was such that we accept only if both states are correct, then the optimum measurement would be a combination of local measurements.}.

\section{Summary and Conclusions}

In this paper we examined minimum-cost measurements in order to obtain useful tools for quantum information and quantum communications. Knowledge of optimal measurements is important for example for bounding the ability of adversaries in cryptographic protocols to forge messages or learn about a secret key. 
We obtained a series of results concerning minimum-cost measurements. In particular, we showed (1) that the minimum-cost measurement remains the same if we add a constant-row cost matrix to the cost matrix, (2) one can bound the minimum cost from above (below) with an element-by-element greater (smaller) cost matrix, (3) one can bound the cost for a 
sum of cost matrices by the sum of the minimum costs for the individual cost matrices in the sum. We also (4) derived a formal mathematical equivalence between minimum-cost measurements for pure states and minimum-error measurements for mixtures of those pure states. Then we focused on the case of symmetric states, where we (5) derived an expression for the square-root measurement (SRM) and the minimum error for pure states  in terms of the eigenvalues of the Gram matrix for the states which takes a surprisingly simple form (Eq. (\ref{min-error-sym})), and (6) showed that when the cost matrix is circulant, symmetric, has negative elements and is negative semidefinite, then the SRM is the minimum-cost measurement. We (7) gave a particular example, where
we obtained lower and upper bounds for the minimum cost of an arbitrary cost matrix. These results lead us to (8) obtain the minimum-error probability for mixed states which are a particular kind of mixtures of pure symmetric states, 
and a method to bound the minimum-error probability for a larger class of mixed states.

Finally we (9) considered sequences of (that is, tensor products of) individual systems, where the global cost is a function of the local costs. We (i) showed that  if this function is linear, then a combination of local minimum-cost measurements is the global minimum-cost measurement, (ii) if the function is convex or concave we obtain bounds (upper/lower) from the local minimum cost measurements.  We moreover (iii) showed that this is not the case for general functions of the local costs, even if the function is monotonic, and pointed out the connection between this, quantum state elimination measurements and the PBR argument regarding the nature of the wave function.

\emph{Acknowledgments. Support by EPSRC grants EP/G009821/1, EP/K022717/1 and an EPSRC Doctoral Fellowship is gratefully acknowledged. PW is also partially supported by COST Action MP1006.}

\appendix

\section{\label{app}Minimum-cost measurements on tensor products of symmetric states}

Consider a minimum-cost measurement on a sequence of individual symmetric states.
We will here show that the first three Helstrom conditions are satisfied by the local SRMs, if the local (individual) states are symmetric, for any global cost matrix that is a function of the sum of the local costs.

We consider tensor product states of symmetric local states, where the local costs are circulant and symmetric, and the global cost is some function of the sum of the local costs. We will prove that the tensor product of local SRMs satisfies the first three Helstrom conditions. However, as expected, the inequality conditions are not in general satisfied. Here we will use the same notation and terminology as in section \ref{Section-tensor-cost}.

\TH
 Assume a global tensor product state of local pure symmetric states, and a global cost matrix that is (any) function of the sum of some local cost matrices. If the local cost matrices are circulant and symmetric, then the first three Helstrom conditions hold for the measurement with measurement which is a combination of local SRMs.
\HT
\proof
We rewrite the minimum-cost measurement for the global system as a minimum-error measurement for newly defined states $\bar\rho_{k_i}=\sum_{k_j}C_{k_i,k_j}\rho_{k_j}$, with the same convention for indices as in 
section \ref{Section-tensor-cost}.
The Helstrom condition for the minimum-error measurement is then 
\EQ{\label{Helstrom_1_tensor}\Pi_{k_1}\left(\sum_{k_1,k_2}C_{k_1,k_3}\rho_{k_3}-\sum_{k_4}C_{k_2,k_4}\rho_{k_4}\right)\Pi_{k_2}=0}
for all global states labelled by 
$k_1,k_2$. By assumption, the cost matrix is of the form
\EQ{C_{k_1,k_2}=f\left(\sum_i C_{k_1(i),k_2(i)}\right).}
We can view the sum of the local cost matrices as a distance of the string $k_1$ from the string $k_2$, and therefore the cost matrix is some function of the distance between the two states. 
The claim is that the tensor product of local SRMs satisfies Eq. (\ref{Helstrom_1_tensor}). The global states corresponding to the SRM are of the form $\ket{\phi_k}=\otimes_i\ket{\phi_{k(i)}}$ and eq. (\ref{Helstrom_1_tensor}) becomes
\EQ{\sum_{k_3,k_4}\left(C_{k_1,k_3}\bra{\phi_{k_1}}\psi_{k_3}\rangle\bra{\psi_{k_3}}\phi_{k_2}\rangle-C_{k_4,k_2}\bra{\phi_{k_1}}\psi_{k_4}\rangle\bra{\psi_{k_4}}\phi_{k_2}\rangle\right)=0.}
To prove that this holds, it is sufficient to show that 
each term in the first sum cancels a term in the second sum, in a way so that the whole sum vanishes. 
We can explicitly show that this is the case. For any given $k_1,k_2,k_3$, choose $k_4$ so that for each element $k_4(i)=k_1(i)+k_2(i)-k_3(i)$, where the addition and subtraction is done for 
the labels of the local symmetric states, and is done modulo $N$. This gives a bijective map between terms in the two sums.
Since the cost matrix is a function of the sum of the local cost matrices, and the local cost matrices are circulant, the total cost matrix is also circulant and therefore
\EQ{C_{k_4,k_2}=C_{k_1+k_2-k_3,k_2}=C_{k_1,k_3},}
where the addition of global indices is understood as element by element addition modulo $N$. 
What remains for the proof is to show that
\EQ{\label{3.1}\bra{\phi_{k_1}}\psi_{k_3}\rangle\bra{\psi_{k_3}}\phi_{k_2}\rangle=\bra{\phi_{k_1}}\psi_{k_4}\rangle\bra{\psi_{k_4}}\phi_{k_2}\rangle}
for the choice of $k_4$ we made above. Note that
\EQ{\bra{\phi_{k_1(i)}}\psi_{k_2(i)}\rangle&=&\bra{\psi_{k_1(i)}}\Phi^{-1/2}\ket{\psi_{k_2(i)}}=\nonumber\\=\bra{\psi_{k_1(i)}}\phi_{k_2(i)}\rangle&=&\bra{\psi_{k_1(i)+l}}\phi_{k_2(i)+l}\rangle,}
i.e. these terms are circulant. The r.h.s. of Eq. (\ref{3.1}) becomes
\EQ{\prod_i \bra{\phi_{k_1(i)}}\psi_{k_4(i)}\rangle\prod_{i'}\bra{\psi_{k_4(i)}}\phi_{k_2(i)}\rangle
&=&\prod_i \bra{\phi_{k_1(i)}}\psi_{k_1(i)+k_2(i)-k_3(i)}\rangle\times\\&&\times\prod_{i'}\bra{\psi_{k_1(i)+k_2(i)-k_3(i)}}\phi_{k_2(i)}\rangle\nonumber \\
&=&\prod_i \bra{\phi_{k_3(i)}}\psi_{k_2(i)}\rangle\prod_{i'}\bra{\psi_{k_1(i)}}\phi_{k_3(i)}\rangle\nonumber\\
&=&\prod_i \bra{\psi_{k_3(i)}}\phi_{k_2(i)}\rangle\prod_{i'}\bra{\phi_{k_1(i)}}\psi_{k_3(i)}\rangle,\nonumber}
using the fact that the local cost matrices are circulant. The last line is equal to the l.h.s. of Eq. (\ref{3.1}) which then shows that eq. (\ref{Helstrom_1_tensor}) holds and completes the proof.\qed

The important thing to note is that we did not need to make any assumptions on the exact form of the global cost function. One can explicitly check that the inequality condition also holds for linear global cost functions, which is expected due to the results of section \ref{Section-tensor-cost}. As we show, it turns out that this condition is often not satisfied, even for certain monotonic functions.

 \end{document}